\documentclass[lettersize,journal]{IEEEtran}
\usepackage{amsmath,amsfonts}
\usepackage{algorithmic}
\usepackage{algorithm}
\usepackage{array}
\usepackage[caption=false,font=normalsize,labelfont=sf,textfont=sf]{subfig}
\usepackage{color}
\usepackage{textcomp}
\usepackage{stfloats}
\usepackage{url}
\usepackage{verbatim}
\usepackage{graphicx}
\usepackage{cite}
\usepackage{amssymb}
\usepackage{bm}
\begin{document}
	
\title{Secrecy Performance Analysis of RIS-aided Communication System with Randomly Flying Eavesdroppers}
\author{Lai Wei, Kezhi Wang,~\IEEEmembership{Member,~IEEE,} Cunhua Pan,~\IEEEmembership{Member,~IEEE,} Maged Elkashlan,~\IEEEmembership{Senior Member,~IEEE}
\thanks{This work of Lai Wei was supported by China Scholarship Council. (\emph{Corresponding author: Kezhi Wang and Cunhua Pan.})}
\thanks{Lai Wei is with Department of Electronics, Peking University, Beijing 100871, China and School of Electronic Engineering and Computer Science, Queen Mary University of London, London E1 4NS, U.K. (e-mail: future1997@pku.edu.cn, lai.wei@qmul.ac.uk).}
\thanks{Kezhi Wang is with Department of Computer and Information Sciences, Northumbria University, NE2 1XE, Newcastle upon Tyne, U.K. (e-mail: kezhi.wang@northumbria.ac.uk).}
\thanks{Cunhua Pan is with the National Mobile Communications Research Laboratory, Southeast University, Nanjing 210096, China (email: cpan@seu.edu.cn).}
\thanks{Maged Elkashlan is with School of Electronic Engineering and Computer Science, Queen Mary University of London, London E1 4NS, U.K. (e-mail: maged.elkashlan@qmul.ac.uk).}}

\markboth{}%
{How to Use the IEEEtran \LaTeX \ Templates}

%\IEEEpubid{}

\maketitle

\begin{abstract}
		 In this letter, we analyze the secrecy performance of a reconfigurable intelligent surface (RIS)-aided communication system with spatially random unmanned aerial vehicles (UAVs) acting as eavesdroppers. \textcolor{black}{We consider the scenarios where the base station (BS) is equipped with single and multiple antennas}. The signal-to-noise ratios (SNRs) of the legitimate user and the eavesdroppers are derived analytically and approximated through a computationally effective method. The ergodic secrecy capacity is approximated and derived in closed-form expressions. Simulation results validate the accuracy of the analytical and approximate expressions and show the security-enhanced effect of the deployment of the RIS.
\end{abstract}
	
	
	\section{Introduction}
	 Unmanned aerial vehicles (UAVs) have been regarded as one of the promising techniques in wireless communication systems. The UAVs could not only serve as flying base stations but also as relays to enhance performance of wireless networks, which help provide broader signal coverage and higher data rates in both military and civilian applications. 
	 
	 However, due to the flexibility of placement and the high mobility of the UAVs, they could be applied to various illegal activities such as launching attacks, conducting surveillance and creating security threats. Moreover, the higher chance of a line-of-sight (LoS) link makes wireless networks more vulnerable to UAV eavesdroppers compared with ground eavesdroppers \cite{b21}. Recently, the physical layer security (PLS) of UAV-enabled wireless communications have attracted extensive research interests. In \cite{b1}, the authors analyzed the secrecy performance of a UAV-to-UAV communication system considering large-scale fading channel. In \cite{b2}, the detection probability of a wireless system with a spatially random UAV eavesdropper was investigated. Furthermore, to enhance the secrecy performance in UAV-enabled wireless systems, several techniques have been proposed, such as jamming\cite{b12}, trajectory design\cite{b13} and artificial noise\cite{b14}, etc. 
	 
	 Among the existing PLS schemes, the reconfigurable intelligent surface (RIS)-aided secure communication has attracted wide attentions \cite{b10}. Through sophisticated phase shift design, RIS could improve legitimate user's received SNR without increasing power budget. In addition, due to the randomness of the wireless channel, the signal reflected by the RIS could not only be added constructively at the legitimate user, but also destructively at the eavesdroppers \cite{b5}. \textcolor{black}{The application of RIS in UAV-enabled wireless system could improve the secrecy performance with lower energy and cost. Nevertheless, the existing literature mainly focused on the scenario where the UAVs are acting as friendly relays or flying BSs to enhance secrecy performance of RIS-aided wireless systems, which inspires us to analyze the case where the UAVs act as eavesdroppers. Moreover, the impact of the randomness of Eve's spatial distribution on RIS-aided wireless systems has not been fully addressed in existing works}.
	 
	 \textcolor{black}{Against the above background, in this work, we study the secrecy performance of a RIS-aided wireless system in the presence of spatially random UAVs acting as eavesdroppers. Firstly, we derive the cumulative distribution functions (CDFs) of the SNRs of the legitimate user and the eavesdroppers in analytical forms where the BS is equipped with single antenna. Next, we approximate the results through a computationally effective method. Then, we consider the scenario where the BS is equipped with multiple antennas. Finally, we carry out Monte-Carlo simulations to validate the accuracy of the analytical and the approximate results}.
	
	\section{System Model}
	\textcolor{black}{We consider a RIS-aided communication system with one BS and one legitimate user (i.e., Bob), where Bob is equipped with a single antenna}. Due to the blockage caused by obstacles such as buildings and trees, the direct link between the BS and Bob is negligibly weak. Therefore, a RIS is applied to assist end-to-end communication. Furthermore, there exists a group of UAV acting as eavesdroppers (i.e., Eves) aiming to eavesdrop the information of the legitimate user.
	
%	\begin{figure}[htbp]
%		\vspace{-1em}
%		\centering
%		\includegraphics[width=0.39\textwidth]{Model.pdf}
%		\caption{System Model}
%		\vspace{-1em}
%	\end{figure}
	
	We assume that the BS is located on the origin of a Cartesian coordinate. \textcolor{black}{For the sake of tractability, we assume that the UAVs are distributed randomly within a hemisphere with radius $R$ surrounding the BS. We further assume that the spatial distribution of the UAVs follows a homogeneous 3-D Poisson point process (PPP) $\mathcal{S}$ with density $\rho_S$}. It is assumed that the RIS is installed on the point $(0,D,0)$ where $D$ denotes the distance between the BS and the RIS. The legitimate user Bob is assumed to be located at a fixed location. Also, the BS-UAV and the RIS-UAV communication links are assumed to follow Rician fading while the BS-RIS and the RIS-Bob links are assumed to follow Rayleigh fading. Throughout this paper, we use $d_B$ to denote the distance between the RIS and Bob and $d_{p,E_k}(p\in \left\{{\rm{BS,RIS}}\right\})$ to denote the distance between position $p$ and the $k$-th UAV. 
	Moreover, we use $h_i$ and $g_i (i=1,2,\cdots N)$ to denote the small-scale fading of BS-RIS and RIS-Bob channel with $i$ denoting the index of RIS element, respectively.
\color{black}	
\section{Secure Performance Analysis in Single Antenna Scenario}
\color{black}
\subsection{Legitimate Link}
	\textcolor{black}{We first consider the case where the BS is equipped with a single antenna}. With the aid of the RIS, the received SNR at the legitimate user Bob could be expressed as
	\begin{equation}
		\gamma_B = \rho {\left( Dd_B \right)^{-\alpha}} {\left| \sum_{i = 1}^{N} g_i h_i e^{j\phi_i} \right|} ^2, 
		%= \rho {\left( Dd_u \right)^{-\alpha}} {\left( \sum_{i = 1}^{N} \xi_i \right)} ^2.
	\end{equation}
	where $\phi_i$ is the phase shift applied by the $i$-th RIS element, $\rho=\frac{P}{\sigma_0^2}$ denotes the normalized transmit power of the BS and \textcolor{black}{$\alpha$ is the pathloss factor for the BS-RIS and the RIS-Bob links}.
	
	Assuming perfect instantaneous CSI is available at the BS, the phase shift that maximizes the Bob's received SNR is given by
	\begin{equation}
		\phi_i^* = -\angle{g_i} - \angle{h_i},
	\end{equation}
	which yields the optimal instantaneous \textcolor{black}{SNR} of Bob as follows
	\begin{equation}
		\gamma_B = \rho {\left( Dd_B \right)^{-\alpha}} {\left( \sum_{i = 1}^{N} \left| g_i\right| \left| h_i \right| \right)} ^2 \triangleq \rho {\left( Dd_B \right)^{-\alpha}} Z. 
		%= \rho {\left( Dd_u \right)^{-\alpha}} {\left( \sum_{i = 1}^{N} \xi_i \right)} ^2.
	\end{equation}
	
	According to \cite{b4}, the random variable $Z = ( \sum_{i = 1}^{N} \left| g_i\right| \left| h_i \right| ) ^2$ could be approximated by a Gamma distributed random variable with shape parameter $\hat{k} = \frac{\mathbb{E}[Z]^2}{\mathbb{E}[Z^2]-\mathbb{E}[Z]^2}$ and scale parameter $\hat{\theta}=\frac{\mathbb{E}[Z^2]-\mathbb{E}[Z]^2}{\mathbb{E}[Z]}$, in which the first and second order moments of random variable $Z_u$ could be calculated as
	\begin{equation}
		\begin{split}
			& \mathbb{E}[Z] = a_1N + a_2N(N-1)\\
			& \mathbb{E}[Z^2] = b_1N(N-1)(N-2)(N-3)\\
			& + b_2N(N-1)(2N-1) + b_3N^2+b_4N,
		\end{split}
	\end{equation}
	where the constants are given by $a_1=1,a_2=\frac{\pi^2}{16},b_1=\frac{\pi^4}{256},b_2=\frac{3\pi^2}{16},b_3=3,b_4=1$\cite{b3}, respectively.
	
	The CDF of Bob's received SNR could then be approximated as
	\begin{equation}
		\label{FgammaB}
		F_{\gamma_B}(x) \approx \frac{1}{\Gamma(\hat{\theta})} \gamma \left( \hat{k},\frac{x}{\rho(Dd_B)^{-\alpha}\hat{\theta}} \right),
	\end{equation}
	where $\gamma(s,z)=\int_{0}^{z}e^{-t}t^{s-1}{\rm{d}}t$ denotes the lower incomplete Gamma function and $\Gamma(z)$ denotes the Gamma function.
\subsection{Eavesdropping Link}
	Considering both the reflected link and the direct link, the received SNR at the $k$-th Eve $E_k$ is given by
	\begin{equation}
		\begin{split}
		&\gamma_{E_k} \triangleq  \rho\left| Z_{E_k} \right|^2\\
		&= \rho  {\left| {\left( Dd_{RIS,E_k} \right)}^{-\frac{\alpha}{2}}\sum_{i=1}^{N} v_{k,i} h_i e^{j\phi_i} + {\left( d_{BS,E_k} \right)}^{-\frac{\alpha}{2}}v_{k,0} \right|}^2,
		\end{split}
	\end{equation} 
	where $v_{k,i}$ denotes the small-scale fading between $E_k$ and the $i$-th RIS element while $v_{k,0}$ denotes the channel coefficient of the direct link between $E_k$ and the BS. For the sake of simplicity, we define the channel vector between $E_k$ and all the RIS elements as follows
	\begin{equation}
		\mathbf{v}_k \triangleq [v_{k,1},\cdots,v_{k,N}]^T
		= \sqrt{\frac{\beta_1}{\beta_1+1}}\mathbf{\bar{v}}_k + \sqrt{\frac{1}{\beta_1+1}}\widetilde{\mathbf{v}}_k.
	\end{equation}
	Similarly, the channel between the BS and $E_k$ can be given by
	\begin{equation}
			v_{k,0} = \sqrt{\frac{\beta_2}{\beta_2+1}}\bar{v}_{k,0} + \sqrt{\frac{1}{\beta_2+1}}\widetilde{v}_{k,0},
	\end{equation}
	where $\beta_1$ and $\beta_2$ denote the Rician factors. 
	
	\textcolor{black}{
		According to \cite{b5}, $\mathbf{\bar{v}}_k$ could be calculated as
		\begin{equation}
			\begin{split}
				& \mathbf{\bar{v}}_k = e^{-j\frac{2\pi d_{RIS,E_k}}{\lambda}}\times[1,\cdots,e^{-j\frac{2\pi(N_r-1)\delta_r\sin{\xi_k}\cos{\psi_k}}{\lambda}}]^T\\
				& \otimes [1,e^{-j\frac{2\pi 	\delta_c\sin{\xi_k}\cos{\psi_k}}{\lambda}},\cdots,e^{-j\frac{2\pi(N_c-1)\delta_c\sin{\xi_k}\cos{\psi_k}}{\lambda}}]^T.
			\end{split}
		\end{equation}
%	$\mathbf{\bar{h}}=e^{-j\frac{2\pi D}{\lambda}}\times[1,e^{-j\frac{2\pi d\cos{\phi_i}}{\lambda}},\cdots,e^{-j\frac{2\pi(N-1)d\cos{\phi_i}}{\lambda}}]^T$,
	where $\lambda=\frac{c}{f}$ is the carrier wavelength, $\delta_r$ and $\delta_c$ are respectively the row and the column spacing of RIS elements, $\xi_k$ and $\psi_k$ depict respectively the azimuth and elevation angles of departure (AoD) from the RIS to $E_k$. Moreover, the symbol $\otimes$ represents the Kronecker product.
	}
 	Similarly, we have $\bar{v}_{k,0} = e^{-j\frac{2\pi d_{BS,E_k}}{\lambda}}$. Furthermore, the NLoS components of $\mathbf{v}_k$ and $v_{k,0}$ could be given as $\mathbf{\widetilde{v}}_k\sim\mathcal{CN}(0,\mathbf{I})$, $\widetilde{v}_{k,0}\sim\mathcal{CN}(0,1)$, respectively.

	Using spherical coordinate transformation, the RIS is located on the point $(D,\pi/2,\pi/2)$, thus the distance between $E_k$ and the RIS could be expressed as 
	\begin{equation}
		d_{RIS,E_k} = \sqrt{r_k^2 + D^2 - 2r_kD \sin{\theta_k}\sin{\varphi_k}},
	\end{equation}
	where $r_k$, $\theta_k$ and $\varphi_k$ denote the radius, the polar angle and the azimuth angle of $E_k$, respectively. Note that $r_k=d_{BS,E_k}$, thus we substitute $r_k$ for $d_{BS,E_k}$ in the following analysis. 

	Since $\gamma_{E_k}$ involves the randomness of large-scale fading, small-scale fading and the sum of a series of random variables, the exact distribution of $\gamma_{E_k}$ is difficult to get, if not impossible. Therefore, we obtain an approximate distribution of $\gamma_{E_k}$ in the following.
		
	To begin with, according to central limit theorem (CLT), the random variable $\sum_{i=1}^{N} v_{k,i} h_i e^{j\phi_i}$ could be approximated by a complex Gaussian random variable with zero mean and variance $N$, namely, $\sum_{i=1}^{N} v_{k,i} h_i e^{j\phi_i}\sim\mathcal{CN}(0,N)$\cite{b10}. %$\bar{X}_{E_k}\triangleq\frac{1}{\sqrt{{N}}}\sum_{i=1}^{N} v_i h_i e^{j\phi_i}$ could be approximated by a complex Gaussian random variable, the mean and variance of which could be calculated as
	%\begin{equation}
		%\begin{split}
			%& \mathbb{E}\left\{X_{E_k}\right\} = \mathbb{E}\left\{v_ih_ie^{j\phi_i}\right\} = \mathbb{E}\left\{v_i\right\}\mathbb{E}\left\{h_i\right\}\mathbb{E}\left\{e^{j\phi_i}\right\} \overset{(a)}{=} 0 \\
			%& \mathbb{V}\left\{X_{E_k}\right\} = \mathbb{E} \left\{ {\left| v_ih_ie^{j\phi_i} \right|}^2 \right\} - \mathbb{E}^2\left\{v_ih_ie^{j\phi_i}\right\} \\
			%& = \mathbb{E} \left\{ {\left| v_i \right|}^2\right\} \mathbb{E} \left\{ {\left| h_i \right|}^2\right\}  \mathbb{E} \left\{ {\left| e^{j\phi_i} \right|}^2\right\} \overset{(b)}{=} 1
		%\end{split}.
	%\end{equation}
	
	Next, because the randomness of $d_{RIS,E_k}$ is difficult to handle, similar to \cite{b6}, we assume that the distance between the BS and $E_k$ is much less than the distance between the BS and the RIS. As a result, the distance between the RIS and $E_k$ could be approximated by $D$, namely, $d_{RIS,E_k} \approx D$. 
	
	Moreover, since $v_{k,0}$ is a Gaussian random variable and the sum of independent Gaussian random variables is still Gaussian distributed, the distribution of $Z_{E_k}$ conditioned on $r_k$ could be approximated by a complex Gaussian random variable, namely,
	$Z_{E_k}|r_k\sim\mathcal{CN}\left(r_k^{-\frac{\alpha}{2}}\sqrt{\frac{\beta_2}{\beta_2+1}}e^{-j\frac{2\pi r_k}{\lambda}},\frac{N}{D^{2\alpha}}+\frac{r_k^{-\alpha}}{\beta_2+1}\right)$. 
	
	Then, $\gamma_{E_k}$ becomes a non-central Chi-square distributed random variable with two degrees-of-freedom, the CDF of which could be given as follows using Marcum-Q function
	\begin{equation}
		F_{\gamma_{E_k}}(x) = 1 - Q_1(\sqrt{2\mu}/\sigma_{E_k} , \sqrt{2x}/\sqrt{\rho}\sigma_{E_k}),
	\end{equation}
	where $\mu = r_k^{-\alpha}\frac{\beta_2}{\beta_2+1}$ and $\sigma_{E_k}^2 = \frac{N}{D^{2\alpha}}+\frac{r_k^{-\alpha}}{\beta_2+1}$.
	
%	Since the exact distribution function of $\gamma_{E_k}=\rho|X|^2$ is hard to obtain, we approximate $\gamma_{E_k}$ as a Gamma-distributed random variable, the shape and scale parameter of which could be approximated as follows
%	\begin{equation}
%		\begin{split}
%			\kappa = \frac{(\mathbb{E}[|X|^2])^2}{\mathbb{V}[|X|^2]} = \frac{(\mu^2+2\sigma^2)^2}{4\sigma^2(\sigma^2+\mu^2)} = \\
%			\nu = \frac{\mathbb{V}(|X|^2)}{\mathbb{E}[|X|^2]} = \frac{4\sigma^2(\sigma^2+\mu^2)}{\mu^2+2\sigma^2}
%		\end{split}
%	\end{equation}
	
	Considering non-colluded eavesdroppers, the CDF of the overall eavesdropping SNR could be derived as
	\begin{equation}
		\begin{split}
			&F_{\gamma_E}(x)=\max_{k\in\mathcal{S}}F_{\gamma_{E_k}}(x)\\ 
			&=\mathbb{E}_{r_k,\theta_k,\varphi_k}\left[\prod_{k\in\mathcal{S}}F_{\gamma_{E_k}}(x|r_k,\theta_k,\varphi_k)\right]\\
			&= \exp \left\{ -\int_{\mathcal{S}} \rho_S \left[ 1 - F_{\gamma_{E_k}}(x|r_k,\theta_k,\varphi_k) \right]{\rm{d}}r_k{\rm{d}}\theta_k{\rm{d}}\varphi_k \right\} \\
			%& = \exp \left\{ %-\int_{0}^{R}\int_{0}^{2\pi}\int_{0}^{\fra%c{\pi}{2}} \rho_S r^2 \sin{\varphi} %\left[ 1 - %F_{\gamma_{E_m}}(x|d_{BS,E_m},\theta_m,\va%rphi_m) %\right]{\rm{d}}r{\rm{d}}\theta{\rm{d}}\var%phi \right\}\\
			%& = \exp \left\{ -\int_{0}^{R}\int_{0}^{2\pi}\int_{0}^{\frac{\pi}{2}} \rho_S r^2 \sin{\varphi}\right.\\ 
			%&\left.Q_1\left(\frac{\sqrt{2r^{-\alpha }\frac{\beta_2}{\beta_2+1}}}{\sqrt{ND^{-2\alpha}+\frac{r^{-\alpha}}{\beta_2+1}}},\frac{\sqrt{2x}}{\sqrt{\rho \left(ND^{-2\alpha}+\frac{r^{-\alpha}}{\beta_2+1}\right)}}\right){\rm{d}}r{\rm{d}}\theta{\rm{d}}\varphi \right\}\\
			& = \exp\left\{ -2\pi\rho_S \int_{0}^{R} r^2 \right.\\
			& \left.Q_1\left(\frac{\sqrt{\frac{2\beta_2}{\beta_2+1}}}{\sqrt{N(\frac{r}{D^2})^{\alpha}+\frac{1}{\beta_2+1}}},\frac{\sqrt{2x}}{\sqrt{\rho(\frac{N}{D^{2\alpha}}+\frac{r^{-\alpha}}{\beta_2+1})}}\right){\rm{d}}r\right\}.
		\end{split}
	\end{equation}
	For the sake of simplicity, we define the following function
	\begin{equation}
		I(x) = \int_{0}^{R} r^2  Q_1(\frac{\sqrt{\frac{2\beta_2}{\beta_2+1}}}{\sqrt{N(\frac{r}{D^2})^{\alpha}+\frac{1}{\beta_2+1}}},\frac{\sqrt{2x}}{\sqrt{\rho(\frac{N}{D^{2\alpha}}+\frac{r^{-\alpha}}{\beta_2+1})}}){\rm{d}}r.
	\end{equation}

	Due to the complicated structure of Marcum-Q function, it is challenging to obtain a closed-form expression of $I(x)$. Therefore, we make further approximations to calculate the function $I(x)$ defined above. 
	
	Recalling that we assume that the distance between the BS and the RIS is much larger than that between the BS and $E_k$, it is reasonable to treat the item $\frac{r}{D^2}$ as a small variable. With the help of the series expansion of Marcum-Q function \cite{b8}, we approximate the Marcum-Q function in $I(x)$ as follows
	\begin{equation}
		\begin{split}
		& Q_1\left(\frac{\sqrt{\frac{2\beta_2}{\beta_2+1}}}{\sqrt{N(\frac{r}{D^2})^{\alpha}+\frac{1}{\beta_2+1}}},\frac{\sqrt{2x}}{\sqrt{\rho(\frac{N}{D^{2\alpha}}+\frac{r^{-\alpha}}{\beta_2+1})}}\right) \\
		& \approx Q_1\left( \sqrt{2\beta_2}, \sqrt{2(\beta_2+1)r^{\alpha}[1-(\beta_2+1)N(\frac{r}{D^2})^{\alpha}]x/\rho} \right)\\
		& \approx \sum_{n=0}^{+\infty}\sum_{k=0}^{n} \frac{e^{-\beta_2}(\beta_2)^n}{n!k!}\exp\left(-(\beta_2+1)r^{\alpha}x/\rho\right)\\
		& \times [1-(\beta_2+1)N(\frac{r}{D^2})^{\alpha}]^{k}\left(\frac{(\beta_2+1)r^{\alpha}x}{\rho}\right)^{k}.
		\end{split}
	\end{equation}
	 %Using the series expansion of Marcum-Q function , we further express the function as 
	%\begin{equation}
	%	\begin{split}
	%	& Q_1\left( \sqrt{2\beta_2}, \frac{\sqrt{2x}}{\sqrt{\rho(ND^{-2\alpha}+\frac{r^{-\alpha}}{\beta_2+1})}}\right)=\sum_{n=0}^{+\infty}\sum_{k=0}^{n} \frac{e^{-\beta_2}(\beta_2)^n}{n!k!} \\ 
	%	& \times \exp\left({-\frac{x}{\rho(ND^{-2\alpha}+\frac{r^{-\alpha}}{\beta_2+1})}}\right)\left(\frac{x}{\rho(ND^{-2\alpha}+\frac{r^{-\alpha}}{\beta_2+1})}\right)^{k}
	%	\end{split}.
	%\end{equation}
	
	 Using binomial expansion and the integral formula \cite[(8.350.1)]{b17}, the function $I(x)$ could be calculated as
	
%	\begin{equation}
%		\begin{split}
%			I_1 & \approx 	%\sum_{n=0}^{+\infty}\sum_{k=0}^{n}e^{-\beta_2}\frac{%(\beta_2)^n}{n!} \\
%			& \frac{1}{\alpha %k!}\left((\beta_2+1)x\right)^{-\frac{3}{\alpha}}\lef%t[\gamma\left(k+\frac{3}{\alpha},(\beta_2+1)xR^{\alp%ha}\right) - %\frac{kN}{x(d_0D)^{\alpha}}\gamma\left(k+\frac{3}{\a%lpha}+1,(\beta_2+1)xR^{\alpha}\right)\right]
%		\end{split}
%	\end{equation}
	
	\begin{equation}
		\label{Ix}
		\begin{split}
			I(x) & \approx 	\sum_{n=0}^{+\infty}\sum_{k=0}^{n}\sum_{l=0}^{k}\frac{e^{-\beta_2}}{\alpha(\beta_2+1)^{\frac{3}{\alpha}}}\frac{(\beta_2)^n}{n!k!}\binom{k}{l}\left(- \frac{N}{D^{2\alpha}} \right)^{l}
			\\ & 
			\left(\frac{x}{\rho}\right)^{-\frac{3}{\alpha}-l}\gamma\left(k+l+\frac{3}{\alpha},(\beta_2+1)\frac{x}{\rho}R^{\alpha} \right).
		\end{split}
	\end{equation}
	Note that an infinite summation is involved in the above result. To reduce the computational complexity as well as preserve approximation accuracy, we truncate the summation up to $\bar{n}$. As shown by our simulation results in the following section, the approximate expressions match their exact values very well.
	
	Then, the CDF of the overall eavesdropping SNR is given by
	\begin{equation}
		\label{FgammaE}
		F_{\gamma_E}(x) = \exp\left( -2\pi\rho_S I(x) \right).
	\end{equation}
%	The CDF of the distance $d_{E_k}$ between the $kth$ UAV eavesdropper and the RIS could be obtained as
%	
%	\begin{equation}
%		F_{d_{E_k}} (x) = \left\{
%		\begin{aligned}
%			\frac{x^3}{R^3} & , & x \leq R \\
%			1 & , & x > R
%		\end{aligned}.
%		\right.
%	\end{equation}
%	
%	According to central limit theorem (CLT), the random variable $X_{E_k} = \sum_{i=1}^{N} v_i h_i e^{j\phi_i}$ could be approximated with a complex Gaussian random variable. With the CDF of $d_{E_k}$ and the approximation of $X_{E_k}$, we obtain the CDF of $\gamma_{E_k}$ as follows
%	\begin{equation}
%		F_{\gamma_{E_k}}(x) = 1 - \frac{3}{\alpha R^3}\left(\frac{x}{N\rho}\right)^{-\frac{3}{\alpha}}\gamma(\frac{3}{\alpha},\frac{x}{N\rho}R^{\alpha}),
%	\end{equation}
%	where $\gamma(\cdot,\cdot)$ denotes the lower incomplete gamma function.
%	
%	To handle the worst-case scenario, we use the maximum received SINR at the noncollusive UAV eavesdroppers to describe the secure performance of the system. On condition that there exists $M$ UAV eavesdroppers, the CDF of 
%	$\gamma_{E_{max}} = \max\{\gamma_{E_1},\gamma_{E_2},\cdots,\gamma_{E_M}\}$ could be calculated as
%	\begin{equation}
%		F_{\gamma_{E_{max}}}(x) = \prod_{k=1}^{M}F_{\gamma_{E_k}}(x) = \left[1 - \frac{3}{\alpha R^3}\left(\frac{x}{N\rho}\right)^{-\frac{3}{\alpha}}\gamma(\frac{3}{\alpha},\frac{x}{N\rho}R^{\alpha}) \right]^{M}
%	\end{equation}

\subsection{Secrecy Capacity Analysis}
	With the aid of $F_{\gamma_B}(x)$ and $F_{\gamma_E}(x)$, the ergodic secrecy capacity could be expressed as follows
	\begin{equation}
		\label{Cs}
		C_s = \frac{1}{\ln2}\left( \int_{0}^{+\infty}\frac{\bar{F}_{\gamma_B}(x)}{1+x}{\rm{d}}x - \int_{0}^{+\infty}\frac{\bar{F}_{\gamma_B}(x)\bar{F}_{\gamma_{E}}(x)}{1+x}{\rm{d}}x \right),
	\end{equation}
	where $\bar{F}_{\gamma_B}(x)$ and $\bar{F}_{\gamma_{E}}(x)$ denote the complementary CDFs of $\gamma_B$ and $\gamma_{E}$, respectively \cite{b2}.
	%which could be expressed as follows
	%\begin{equation}
	%	\begin{split}
	%		& \bar{F}_{\gamma_U}(x) = \frac{\Gamma(\hat{k},\frac{x}{\rho(Dd_u)^{\alpha}\hat{\theta}})}{\Gamma(\hat{k})}\\
	%		& \bar{F}_{\gamma_{E_{max}}}(x) = 1 - e^{-2\pi\rho_SI(x)}	
	%	\end{split}.
	%\end{equation}
	
	By substituting (\ref{FgammaB}) and (\ref{FgammaE}) into (\ref{Cs}), using the representations of elementary functions in terms of Meijer’s G-Function \cite[(8.4.2.5), (8.4.16.2)]{b18}\cite[Eq. (21)]{b19}
	% $t = \frac{4}{\pi}\arctan(x) - 1$ and $x = \tan(\frac{\pi}{4}(t+1))\triangleq\nu(t)$
	and invoking the Gauss-Chebyshev quadrature approximation \cite{b15}, the ergodic secrecy capacity could be expressed as
	%\begin{equation}
	%	\begin{split}
	%		& \bar{C}_s = \frac{1}{\ln2}\left( 	\frac{G^{3,1}_{2,3}\left[\frac{1}{\rho(Dd_u)^{-\alpha}\hat{\theta}}\left.\right|\substack{0,1\\0,\hat{k},0}\right]}{\Gamma(\hat{k})} \right. \\
	%		& - \left. \frac{\pi}{4}\int_{-1}^{1}\frac{\sec^2(\frac{\pi}{4}(t+1))}{1+\nu(t)}\left[ 1 - e^{-2\pi\rho_S I(\nu(t))} \right]\frac{\Gamma(\hat{k},\frac{\nu(t)}{\rho(Dd_u)^{\alpha}\hat{\theta}})}{\Gamma(\hat{k})}{\rm{d}}t \right)
	%	\end{split}.
	%\end{equation}
	%\begin{equation}
	%	\begin{split}
	%		\bar{C}_{Loss}&  = \frac{\pi^2}{4W\ln 2}\sum_{j=1}^{W}\frac{\sqrt{1-t_j^2}}{1+\nu(t_j)}\sec^2(\frac{\pi}{4}(t_j+1))\\
	%		& \left[ 1 - e^{-2\pi\rho_S I(\nu(t_j))} \right]\frac{\Gamma(\hat{k},\frac{\nu(t_j)}{\rho(Dd_u)^{\alpha}\hat{\theta}})}{\Gamma(\hat{k})}
	%	\end{split}.
	%\end{equation}
	\begin{equation}
		\label{Cs_GC}
		\begin{split}
			& C_s \approx \frac{1}{\ln2}\left(  \frac{G^{3,1}_{2,3}\left[\frac{1}{\rho(Dd_B)^{-\alpha}\hat{\theta}}\left.\right|\substack{0,1\\0,\hat{k},0}\right]}{\Gamma(\hat{k})} - \frac{\pi^2}{4W}\sum_{j=1}^{W}\frac{\sqrt{1-t_j^2}}{1+\nu(t_j)} \right.\\ 
			& \left. \sec^2(\frac{\pi}{4}(t_j+1))\left[ 1 - e^{-2\pi\rho_S I(\nu(t_j))} \right]\frac{\Gamma(\hat{k},\frac{\nu(t_j)}{\rho(Dd_B)^{-\alpha}\hat{\theta}})}{\Gamma(\hat{k})} \right),
		\end{split}
	\end{equation}
	where $t_j = \cos(\frac{2j-1}{2W}\pi)$ and $\nu(t_j)=\tan(\frac{\pi}{4}(t_j+1))$.
	
\color{black}
\section{Secrecy Performance Analysis in Multiple Antennas Scenario}
In this section, we consider the scenario where the BS is equipped with multiple antennas. We assume that the number of antenna elements is $K$ and the BS's beamforming vector is $\bf{w}$. The received SNR of Bob could then be calculated as
\begin{equation}
	\gamma_B = \rho(Dd_B)^{-\alpha}\left|{\bf{g}}^H\Phi{\bf{H}}{\bf{w}}\right|^2,
\end{equation}
where ${\bf{g}}_{N\times 1}$, $\Phi = {\rm{diag}}\{e^{j\phi_1},\cdots,e^{j\phi_N}\}$ and ${\bf{H}}_{N\times K}$ denote the RIS-Bob channel, the phase shift matrix of the RIS and the BS-RIS channel, respectively.

The received SNR of $E_k$ could be expressed as
\begin{equation}
		\gamma_{E_k} \approx \rho  {\left| D^{-\alpha}{\bf{v}}_k^H\Phi{\bf{H}}{\bf{w}} + r_k ^{-\frac{\alpha}{2}}{\bf{u}}^H_{k}{\bf{w}} \right|}^2 \triangleq \rho\left| Z_{E_k} \right|^2,
\end{equation}
where ${\bf{v}}_k$ and ${\bf{u}}_{k}$ denote the small scale fading of the RIS-UAV link and the BS-UAV link, respectively. Since it is quite challenging to derive the optimal phase shift design and the exact distribution of $\gamma_B$ and $\gamma_{E_k}$ in general cases, we consider a special case where the phase shifts of RIS are randomly distributed and the BS leverages maximum ratio transmission (MRT) beamforming scheme, namely, we assume that $\phi_i$ is randomly distributed and ${\bf{w}}=\frac{({\bf{g}}^H\Phi{\bf{H}})^H}{||{\bf{g}}^H\Phi{\bf{H}}||}$ with $||\cdot||$ denoting the $\textit{l}^2$-norm of a vector. Under such assumptions, the received SNR of Bob and $E_k$ could be given as
\begin{equation}
	\gamma_B = \rho \sum_{j=1}^{K}\left| \sum_{i=1}^{N}g_i^{*} h_{i,j} e^{j\phi_i} \right|^2,
\end{equation}
and
\begin{equation}
	\gamma_{E_k} = \rho \left|\sum_{j=1}^{K} \left(D^{-\alpha}\sum_{i=1}^{N}v_{k,i}^{*}h_{i,j}e^{j\phi_i}+r_k^{-\frac{\alpha}{2}}u_{k,i}^{*}\right){\rm{w}}_j \right|^2,
\end{equation}
respectively, where $h_{i,j}$ denotes the $(i,j)$-th element of $\bf{H}$, $v_{k,i}$ and $u_{k,i}$ denotes the $i$-th element of ${\bf{v}}_k$ and ${\bf{u}}_k$, ${\rm{w}}_j$ denotes the $j$-th element of $\bf{w}$, and $(\cdot)^{*}$ denotes the conjugate of a complex number. 

Due to the fact that $\mathbb{E}(||{\bf{w}}||^2)=\sum_{j=1}^{k}\mathbb{E}(||\rm{w}_j||^2)=1$, we have $\mathbb{E}(Z_{E,k})=0$ and $\mathbb{E}(|Z_{E,k}|^2) = r_k^{-\alpha}+\frac{N}{D^{2\alpha}}$. Therefore, when the number of antennas is large, according to CLT, the received SNR of the $k$-th eavesdropper could be approximated as a complex Gaussian random variable with zero mean and variance $r_k^{-\alpha}+\frac{N}{D^{2\alpha}}$, namely, $Z_{E_k}|r_k\sim\mathcal{CN}\left(0,r_k^{-\alpha}+\frac{N}{D^{2\alpha}}\right)$. Consequently, the unconditional SNR of the eavesdroppers could be approximated as
\begin{equation}
	F_{\gamma_E}(x) \approx \exp \left\{ -2\pi\rho_S \int_{0}^{R}r^2e^{-\frac{x}{\rho\left(\frac{N}{D^{2\alpha}}+r_k^{-\alpha}\right)}} {\rm{d}}r \right\},
\end{equation}
defining $I(x)\triangleq \int_{0}^{R}r^2e^{-\frac{x}{\rho\left(\frac{N}{D^{2\alpha}}+r_k^{-\alpha}\right)}} {\rm{d}}r$ and similar to the derivation of (\ref{Ix}), we obtain
\begin{equation}
	\begin{split}
	& I(x) \approx \frac{1}{\alpha}\left(\frac{x}{\rho}\right)^{-\frac{3}{\alpha}}\gamma\left(\frac{3}{\alpha},\frac{x}{\rho\left(R^{-\alpha}+\frac{N}{D^{2\alpha}}\right)}\right) + \frac{3}{\alpha}\left(\frac{3}{\alpha}+1\right) \\
	& \times \frac{N}{D^{2\alpha}}\left(\frac{x}{\rho}\right)^{-\frac{3}{\alpha}-1}\gamma\left(\frac{3}{\alpha}+1,\frac{x}{\rho\left(R^{-\alpha}+\frac{N}{D^{2\alpha}}\right)}\right).
	\end{split}
\end{equation}

Similarly, the CDF of Bob's received SNR could be approximated as 
\begin{equation}
	F_{\gamma_B}(x) \approx \frac{1}{\Gamma(K)} \gamma \left( K,\frac{x}{N\rho(Dd_B)^{-\alpha}} \right).
\end{equation}

Finally, leveraging Gauss-Chebyshev quadrature again, the ergodic secrecy capacity in multiple antenna scenario could be calculated in a similar way to (\ref{Cs_GC}).
%\begin{equation}
%	\begin{split}
%		& C_s \approx \frac{1}{\ln2}\left(  \frac{G^{3,1}_{2,3}\left[\frac{1}{N\rho(Dd_B)^{-\alpha}}\left.\right|\substack{0,1\\0,K,0}\right]}{\Gamma(K)} - \frac{\pi^2}{4W}\sum_{j=1}^{W}\frac{\sqrt{1-t_j^2}}{1+\nu(t_j)} \right.\\ 
%		& \left. \sec^2(\frac{\pi}{4}(t_j+1))\left[ 1 - e^{-2\pi\rho_S I(\nu(t_j))} \right]\frac{\Gamma(K,\frac{\nu(t_j)}{N\rho(Dd_B)^{-\alpha}})}{\Gamma(K)} \right),
%	\end{split}
%\end{equation}
%with $t_j = \cos(\frac{2j-1}{2W}\pi)$ and $\nu(t_j)=\tan(\frac{\pi}{4}(t_j+1))$, respectively.

\color{black}
	\section{Simulation Results}
	In this section, we validate the analytical and approximate results through Monte-Carlo simulations. Unless otherwise stated, the parameters are set as follows: $\rho = 20$dB, $N = 100$, \textcolor{black}{$K=200$}, $R = 50$m, $D = 100$m, $d_B = 10$m, $\alpha = 2$, $\rho_S = 10^{-4}$, $\beta_1 = \beta_2 = 3dB$, $f = 2.4$GHz and $\delta = \lambda/4$. Moreover, the infinite summations of the approximated expressions are truncated to $\bar{n} = 10$ and the Gauss-Chebyshev parameter $W$ is set to $W=20$ to reduce computational complexity as well as preserve approximation accuracy.
	
	In Fig.\;1, the approximate and empirical CDFs of the legitimate user and Eves under different average transmit SNRs are presented. 
	Firstly, it could be readily found that the Gamma distribution approximation of the legitimate SNR is close to its exact CDF, which validates the effectiveness of the proposed Gamma distribution approximate procedure. Secondly, the approximation of Eve's received SNR matches the simulation results in various scenarios, which demonstrates the accuracy of our mathematical derivations. \textcolor{black}{Furthermore, when the BS is equipped with multiple antennas, the SNR of Bob increases significantly while the SNR of Eve changes slightly, which demonstrates that the integration of multiple antennas and RIS will be a promising technique to enhance the secrecy performance of wireless systems}. One can also figure out that the approximation of Eve's SNR is much more computationally effective than directly calculating numerical integration, since the infinite summation of the approximation is truncated to tens of times instead of calculating an infinite integral. Actually, the effectiveness of the truncated summation approach could be validated by the mathematical structure of Marcum-Q function according to \cite{b9}. Specifically, a Marcum-Q function could be expressed as a weighted summation of an infinite number of lower incomplete Gamma functions. Due to the monotonically decreasing property of lower incomplete Gamma function $\gamma(n,z)$ with respect to $n$ and the rapid growth speed of $n!$ as $n$ increases, it would be reasonable to neglect the higher order items which contribute a little in the infinite summation, thus an accurate approximation to Marcum-Q function could be obtained henceforth.
	
	\textcolor{black}{In Fig.\;2, we plot the secrecy capacity versus average transmit SNR in single antenna scenario. As observed from Fig.\;2}, the secrecy capacity increases with the transmit power, which demonstrates that the deployment of a RIS greatly enhances the legitimate channel while slightly strengthens the eavesdropping channel. Moreover, the slope of the secrecy capacity versus $\rho$ increases as the number of RIS elements increases, which demonstrates the SNR-enhanced effect of the RIS and motivates the deployment of a larger scale RIS to improve the secrecy performance.
	
	\textcolor{black}{Fig.\;3 depicts the secrecy capacity versus the number of antenna elements $K$. It is shown that the secrecy capacity increases with $K$, which demonstrates that the integration of multiple antennas and RIS is a potential technique in enhancing physical layer security of wireless system}.
	
	Fig.\;4 characterizes the secrecy capacity versus the maximum distance between Eves and the BS. As expected, the secrecy capacity increases as the UAVs gradually moves further from the BS, which illustrates that a setup of a protected zone around the BS to surveil Eves would improve the secrecy performance and alleviate information leakage. Nevertheless, Fig.\;4 presents that the secrecy capacity increases slightly versus $R$ but increases significantly as the number of RIS elements increases, which demonstrates that the deployment of RIS would be a beneficial method to enhance secure information transmissions with lower cost and higher energy efficiency.
	
	\begin{figure}[p]
		\vspace{-1em}
		\centering %图片居中
		\includegraphics[width=0.35\textwidth]{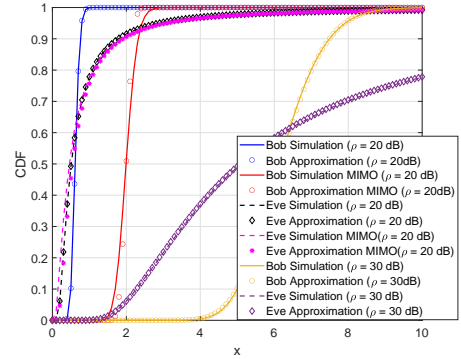} 
		\caption{Approximate and empirical CDFs of legitimate user and UAV Eves.} %最终文档中希望显示的图片标题
		\label{Fig.1} %用于文内引用的标签
		\vspace{-1em} 
	\end{figure}
	
	\begin{figure}[p]
		\vspace{-1em}
		\centering %图片居中
		\includegraphics[width=0.35\textwidth]{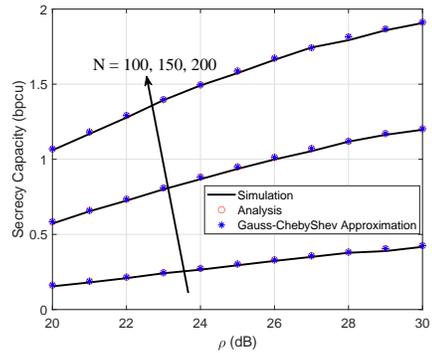} 
		\caption{Ergodic secrecy capacity versus transmit power in single antenna scenario.} %最终文档中希望显示的图片标题
		\label{Fig.2} %用于文内引用的标签
		\vspace{-1em}
	\end{figure}
	
	\begin{figure}[p]
		\vspace{-1em}
		\centering %图片居中
		\includegraphics[width=0.35\textwidth]{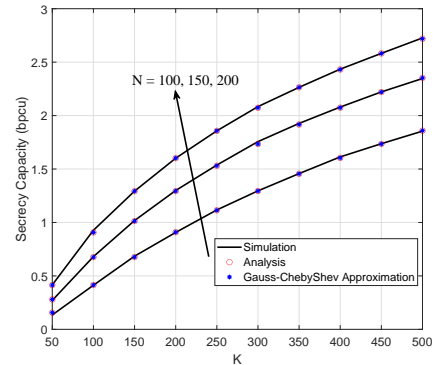} 
		\caption{Ergodic secrecy capacity versus number of antenna elements in multiple antenna scenario.} %最终文档中希望显示的图片标题
		\label{Fig.3} %用于文内引用的标签
		\vspace{-1em}
	\end{figure}
	
	\begin{figure}[p]
		\vspace{-1em}
		\centering %图片居中
		\includegraphics[width=0.35\textwidth]{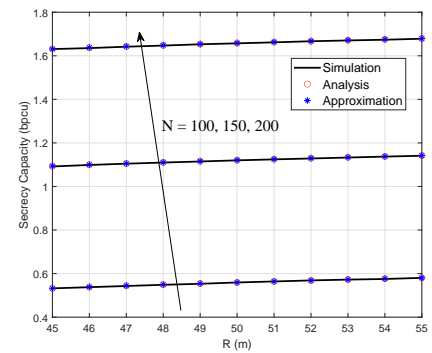} 
		\caption{Ergodic secrecy capacity versus the maximum radius of Eves' distribution area.} %最终文档中希望显示的图片标题
		\label{Fig.4} %用于文内引用的标签
%		\vspace{-1em}
	\end{figure}
	
	\section{Conclusion}
	In this letter, we study the secrecy performance of a RIS-aided communication system with spatially random UAVs acting as eavesdroppers. We derive the analytical expressions for the SNRs of the legitimate user and Eves and approximate the results in a computational effective method, considering both single and multiple antennas scenarios using stochastic geometry. Finally, we validate our results through Monte-Carlo simulations. \textcolor{black}{We have found that the deployment of a larger scale RIS would effectively and economically secure the information transmission. The integration of multiple antennas with RIS and the setup of a protected zone would further enhance the secrecy performance of wireless system}.

	%\newpage

	%\vfill
	
\end{document}